\documentclass[pra,aps,showpacs,twocolumn,floatfix]{revtex4}
\usepackage{epsf}
\usepackage{amsmath}
\usepackage[dvips]{color}
\newcommand {\be}{\begin{equation}}
\newcommand {\ee}{\end{equation}}
\newcommand {\bea}{\begin{eqnarray}}
\newcommand {\eea}{\end{eqnarray}}

\newcommand {\dd} {\mathrm{d}}
\newcommand {\sn} {\mathrm{sn}}

\begin{document}


\title{Period doubling, two-color lattices, and the growth of swallowtails in Bose-Einstein condensates}
\author{B. T. Seaman, L. D. Carr, and M. J. Holland}
\affiliation{JILA, National
Institute of Standards and Technology and Department of Physics,
\\University of Colorado, Boulder, CO 80309-0440}
\date{\today}


\begin{abstract} The band structure of a Bose-Einstein condensate
is studied for lattice traps of sinusoidal, Jacobi elliptic, and
Kronig-Penney form.  It is demonstrated that the physical
properties of the system are independent of the choice of lattice.
The Kronig-Penney potential, which admits a full exact solution in
closed analytical form, is then used to understand in a novel way
the swallowtails, or loops, that form in the band structure. Their
appearance can be explained by adiabatically tuning a second
lattice with half the period. Such a two-color lattice, which can
be easily realized in experiments, has intriguing new physical
properties.  For instance, swallowtails appear even for weak
nonlinearity, which is the experimental regime.  We determine the
stability properties of this system and relate them to current
experiments.
\end{abstract}


\maketitle

\section{Introduction} \label{sec:Introduction}

There has been great interest in the study of Bose-Einstein
condensates (BEC's) from both the experimental and theoretical
perspective since they were first created in
1995~\cite{Anderson1995,Davis1995,Bradley1995}. In particular, the
examination of BEC's in periodic potentials in the superfluid
phase has yielded many intriguing phenomena, such as gap
solitons~\cite{Eiermann2004,Oberthaler2004}, the appearance of
swallowtails or loops in the band structure, pulsed atom lasers
and demonstrations of their phase coherence~\cite{Anderson1998,
Hagley1999}, and matter-wave diffraction~\cite{Ovchinnikov1999}.
BEC's in periodic potentials, unlike other solid state systems,
have the advantage that the lattice geometry and interatomic
interactions are highly
controllable~\cite{Roberts1998,Inouye1998}. We examine the mean
field Bloch states of a BEC in a one-color and two-color potential
for arbitrary interaction and potential strengths. The full
nonlinear band structure of the system is then determined for a
two-color potential.  We show that period-doubled states in the
usual one-color lattice are directly connected with Bloch waves in
the two-color case.  Period-doubled states have a periodicity
which is twice that of the underlying lattice.  This allows for a
novel interpretation of swallowtails, a key physical property of
the band structure of a BEC. The study of period-doubled states
has been important in other systems, for instance in
optics~\cite{Sacher1992,Simpson1994}, because it offers
experimental access to the period-doubling route to chaos.

The band structure of a BEC in an optical lattice has previously
been studied both analytically and numerically.  An array of
techniques have been developed.  The sinusoidal potential, which
is the experimental case, has been investigated by expanding the
wavefunction in a Fourier series and minimizing the
energy~\cite{Wu2001,Diakonov2002,Wu2002,Machholm2003,Machholm2004}.
Two other potentials have been studied in detail.  A Jacobi
elliptic potential, which asymptotically approaches the sinusoidal
case, has the advantage that it emits a useful class of exact
solutions~\cite{Bronski2001,Bronski2001_2,Bronski2001_3}.  On the
other hand, the Kronig-Penney potential can be solved completely
and exactly, as we have shown in our previous
work~\cite{Seaman2005_2,Seaman2005,Li2004}.  We compare these
three models with the aid of the Bloch wave representation.  For
the Jacobi elliptic potential this requires a reinterpretation of
the original result~\cite{Bronski2001}. In addition, we explain
the longstanding open problem of why certain density offsets in
this class of solutions lead to instability.  We find that the
form of the potential has no effect on the superfluid and other
physical properties of the system.  Of the closed-form analytical
methods available for the three potentials, only the Kronig-Penney
potential admits a full description of the band structure.

The two-color lattice has already received some attention in the
literature.  Roth and Burnett showed that when the period of the
second lattice is much longer than the first, so that it forms an
envelope, new quantum phases are introduced into the quantum
problem~\cite{Roth2003}.  We will consider the superfluid phase
only, which is obtained when the lattice height is on the order of
or smaller than the chemical potential.  Very recently, Louis {\it
et al.} have studied gap solitons in a two-color lattice for which
the second lattice has half the period of the
first~\cite{Louis2005}. We study the same potential, but for Bloch
waves, rather than gap solitons.  Bloch waves have the same period
as the lattice, whereas gap solitons are envelope solutions. Using
the Kronig-Penney potential, we link the period-doubled states of
the one-color lattice to Bloch waves in the two-color case.
Adiabatically turning off one of the frequencies in a two-color
lattice, one can clearly observe the origin of swallowtails.
Unlike previous explanations of this intriguing property of
nonlinear periodic systems~\cite{Mueller2002}, ours encompasses
both repulsive and attractive BEC's.

Experiments on BEC's in optical lattice potentials proceed as
follows.  Alkali-metal bosonic atoms are cooled to the quantum
degenerate regime. The interference pattern of two counter
propagating lasers is used to create a sinusoidal potential shift
caused by the ac Stark effect induced by the dipole interaction
with the laser field on the atoms' center of mass
motion~\cite{Meystre1999,Denschlag2002}.  A small frequency
detuning of one of the lasers allows for examination of the
different quasimomentum states and their stability properties by
creating a traveling wave interference pattern moving at the
velocity $v=(\lambda/2)\delta\nu$, where $\lambda$ is the
wavelength of the first beam and $\delta\nu$ is the
detuning~\cite{Fallani2003,Peil2003,Fallani2004}. The two-color
lattice can then be created by the superposition of two lasers
with frequencies which differ by a factor of two. This can be
achieved using second harmonic generation with nonlinear crystals.
Complex lattice configurations with several lasers have already
been performed experimentally~\cite{Guidoni1997}.  We present a
dynamical study of key observables for the two-color lattice, such
as the instability time, that can be experimentally investigated
with minor changes to current apparatus. There have been many
theoretical studies of stability properties of BEC's in one-color
lattices
(see~\cite{Bronski2001,Bronski2001_2,Bronski2001_3,Sukhorukov2001,Hilligsoe2002,Kevrekidis2003_2,Louis2003},
to name a few), though not of the two-color case.  We take
advantage of the analytical simplicity of the Kronig-Penney
potential in order to correctly seed our numerical studies of the
mean field.

Thus, the three main objectives of this article are the following.
First, in Secs.~\ref{sec:NLS} and~\ref{sec:Review}, the
qualitative properties of a condensate in a periodic potential are
shown to be independent of the specific form of the potential.
This justifies the use of a Kronig-Penney potential for the rest
of our study. Second, in Sec.~\ref{sec:TwoColorLattice}, the
nonlinear band structure of the two-color lattice is described
fully and analytically. This is used to interpret the appearance
of swallowtails in the one-color lattice in a novel way.  Third,
in Sec.~\ref{sec:Stability}, numerical simulations are used to
make a connection between the properties of the two-color lattice
and current experiments on BEC's in one-color lattices. Finally,
concluding remarks are made in Sec.~\ref{sec:Conclusion}.

\section{The Nonlinear Schr\"odinger Equation and Bloch Waves}
\label{sec:NLS}

The stationary states of the mean field of a BEC in an external
periodic potential $V(x)$ are governed by the nonlinear
Schr\"odinger equation~\cite{Olshanii1998,Carr2000,Salasnich2002},
also called the Gross-Pitaevskii
equation~\cite{Gross1961,Pitaevskii1961}, \be
-\frac{1}{2}\Psi_{xx}+g|\Psi|^2\Psi + V(x)\Psi = \mu \Psi\, ,
\label{eqn:NLS} \ee where $\mu$ is the eigenvalue and $g$
characterizes the effective quasi-one-dimensional two-body atomic
interaction. A quasi-one dimensional condensate can be created by
making an elongated trap that is tightly bound in the transverse
directions. In Eq. (\ref{eqn:NLS}), length has been scaled by the
lattice spacing $d$ and energy has been scaled by $2 E_0/\pi^2$
where \be E_0 \equiv \frac{\hbar^2 \pi^2}{2 m d^2} \ee is the
recoil energy, or kinetic energy of a particle at the edge of the
first Brillouin zone, and $m$ is the particle's mass. With this
scaling, the renormalized 1D coupling is $g \equiv 2 a_s \omega m
d/\hbar$, where harmonic confinement in the transverse directions
has been assumed with frequency $\omega$, and $a_s$ is the two
body $s$-wave scattering length.

The wavefunction $\Psi$ is given by \be \Psi=\sqrt{\rho(x)}\exp[i
\phi(x)-i\mu t]\, ,  \label{eqn:Psi}\ee where $\rho(x)$ is the
linear density of the condensate and the local superfluid velocity
is given by $v=\partial\phi/\partial x$. Substituting
Eq.~(\ref{eqn:Psi}) into Eq.~(\ref{eqn:NLS}), one finds that the
phase is related to the density by \be \phi(x) = \alpha
\int_0^x\frac{\dd x'}{\rho(x')}\, , \label{eqn:phase}\ee where
$\alpha$ is a real constant of integration.  Note that we have
chosen $\phi(0)=0$.

When describing the band structure of a system, one constructs the
solutions in the usual Bloch form, \be \Psi(x) = e^{iqx} f_q(x)\,
, \ee where $q$ is the quasimomentum and $f_q(x)=f_q(x+1)$, i.e.,
has the same period as the lattice. The band structure $E(q)$ can
be determined from the relationship between the quasimomentum and
the energy per particle in terms of the density and phase of the
solution: \bea q &=& \alpha\int_0^1
\frac{\dd x'}{\rho(x')} \, , \label{eqn:quasimomentum} \\
\frac{E}{n} &=& \frac{1}{n}\int_0^1 \dd
x'(\frac{1}{2}|\partial_{x'}\Psi|^2+\frac{g}{2}|\Psi|^4+V(x')|\Psi|^2)\,
, \label{eqn:energy} \eea where $E/n$ is the energy per atom, $q$
is the quasimomentum, and $n$ is the number of atoms per lattice
site.  The latter is given by the normalization \be n=\int_0^1\dd
x'\rho(x')\, . \ee The quasimomentum is the phase jump between
adjacent sites of the lattice and is proportional to the mean
superfluid velocity of the system.  We scale the quasimomentum by
\be q_0\equiv\pi/d\, , \ee which is the quasimomentum at the edge
of the Brillouin zone.

The band structure is modified from the well-known solutions of
the linear Schr\"odinger equation with a periodic potential. There
are two physical regimes~\cite{Seaman2005_2}. For $|g|n \le V_0$,
which we term the regime of \emph{weak nonlinearity}, the linear
band structure is simply perturbed up or down, depending on the
sign of $g$. For $|g|n \gg V_0$, which corresponds to the regime
of \emph{strong nonlinearity}, the bands wrap back around on
themselves to form loop structures, or swallowtails.  These
swallowtails have previously been described in terms of the
superfluid screening properties of the
condensate~\cite{Mueller2002}.  This explanation is only valid for
repulsive condensates, despite the fact that swallowtails also
appear in the attractive case.  This motivates the need for
another explanation, which we provide in
Sec.~\ref{sec:TwoColorLattice}.

There have been many methods used to determine the nonlinear band
structure of BEC's in periodic structures.  In the following
section, three analytical methods are described and evaluated. In
addition to these analytical approaches, many numerical methods
have been developed.  These include Hamiltonian perturbation
theory~\cite{Porter2004_2} and the study of accelerating
lattices~\cite{Wu2003}.  We consider only analytical methods of
obtaining stationary states.

\section{Analytical Methods in Nonlinear Band Theory} \label{sec:Review}

Of the many methods used in nonlinear band theory, analytical
methods allow for greater flexibility in describing solution
types, as they result in general expressions which describe all
parameter regimes simultaneously.  In this section, we compare
three methods that use different periodic potentials and interpret
their solutions in the Bloch wave representation. It is found that
they all produce the same band structure for the lowest band.
Hence, the exact form of the potential is unimportant.
Nevertheless, only the method based on the Kronig-Penney potential
is able to describe all bands analytically since an exact solution
method exists for piecewise constant
potentials~\cite{Carr2000,Carr2000_2,Seaman2005_2}.


\subsection{Solution by Cancellation} \label{subsec:Bronski}

The first analytical method for obtaining solutions to the NLS
with a periodic potential in the context of a BEC was given by
Bronski, Carr, Deconinck, and Kutz~\cite{Bronski2001} for a Jacobi
elliptic potential of form \be V(x)=-V_0\{1-2\, \sn^2[2 K(k)
x,k]\}\, , \ee where $\sn$ is one of the Jacobi elliptic
functions~\cite{Abramowitz1964}. The Jacobi elliptic functions are
generalized periodic functions characterized by an elliptic
parameter $k \in [0,1]$. They approach circular and hyperbolic
trigonometric functions as $k \to 0$ and $k \to 1$, respectively.
For $k$ not exponentially close to unity, the potential is similar
to the sinusoidal case.  As $k$ approaches unity, the period of
the lattice becomes much greater than the width of the localized
variations of the potential, thereby approaching a lattice of
delta functions, i.e., a Kronig-Penney potential. The period of
the lattice is $2 K(k) \in [\pi,\infty]$ where $K(k)$ is a
complete elliptic integral of the first
kind~\cite{Abramowitz1964}.

Bronski {\it et al.} were able to show that it is possible to
choose a suitable ansatz for the density such that the nonlinear
term cancels the potential term, thereby leaving a system that is
effectively free. They found a special class of solutions with a
period equal to that of the lattice, where the density was assumed
to be of the form \be \rho(x) = A \,\sn^2(b x,k)+B\, .
\label{eqn:bronski1}\ee The following conditions must then be met:
\bea A &=& \frac{b^2 k^2+2 V_0}{g}\, , \\ \alpha^2 &=&
B(B+A)(b^2+B g-\frac{2 B V_0}{A})\, , \\ \mu &=& \frac{1}{2}(b^2+A
g+3 B g)-\frac{B V_0}{A} \, .  \label{eqn:bronski4}\eea

Machholm {\it et al}.~\cite{Machholm2003} showed that this
solution corresponds to the edge of the Brillouin zone for the
exactly sinusoidal case, $k=0$. If the parameters of the
potential, $V_0$, $k$, and $b$, and the interaction strength $g n$
are set, the wavefunction is completely determined without any
free parameters. Nevertheless, the solutions are of more general
use.  In fact, although not described in this way by the original
authors, this family of solutions can be used to map out the whole
lowest energy band.  It is possible to change the elliptic
parameter $k$ to get a spectrum of solutions. Since the Jacobi
elliptic functions closely approximate the trigonometric functions
for all $k$ except exponentially close to unity, there is a wide
range of values for $k$ where the potential is approximately
sinusoidal. In addition, the sign of the potential coefficient
$V_0$ does not significantly change the form of the potential and
so this is another parameter that can be changed to determine the
energy bands.  Therefore, when the elliptic parameter is varied,
and the solutions determined via
Eqs.~(\ref{eqn:bronski1})-(\ref{eqn:bronski4}), the complete
lowest energy band can be analytically determined for arbitrary
interaction strength.

\begin{figure*} [tb]
\begin{center}
\epsfxsize=18cm \leavevmode \epsfbox{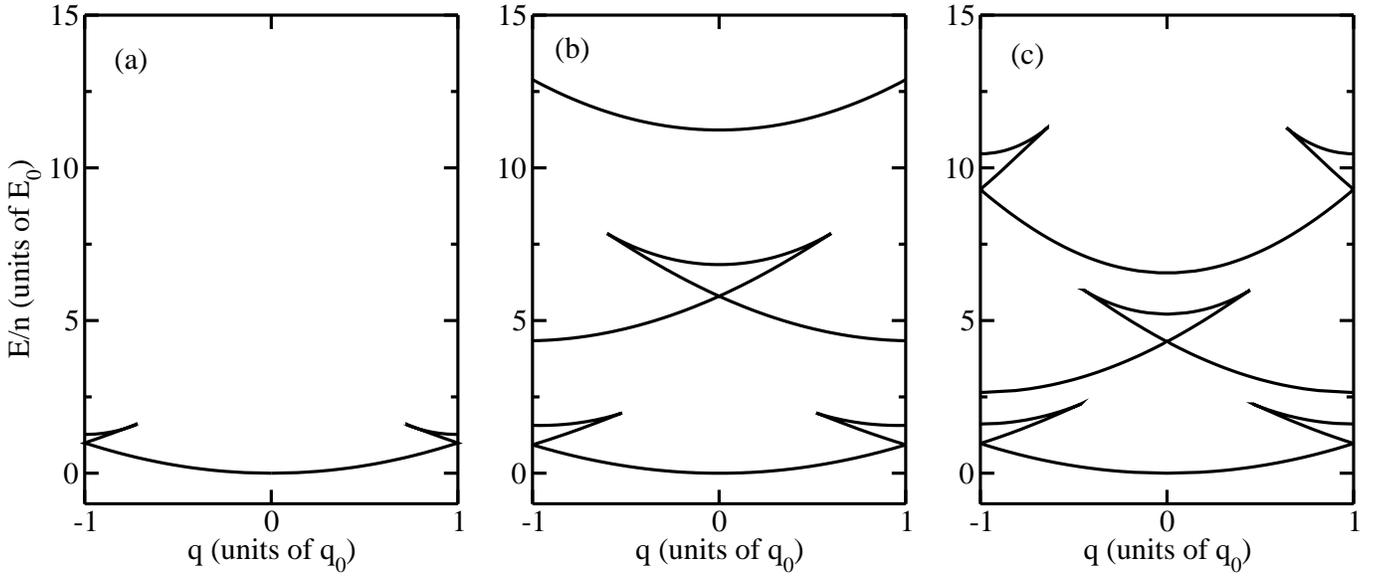}
\caption{Band structure for a strongly repulsive condensate,
$gn=10$, in a periodic potential.  Shown are results obtained via
the analytical methods of (a) Bronski {\it et
al.}~\cite{Bronski2001} with a Jacobi elliptic potential
($|V_0|=1$), (b) Machholm {\it et al.}~\cite{Machholm2003} with a
sinusoidal potential ($V_0=4$), and (c) Seaman {\it et
al.}~\cite{Seaman2005_2} with a Kronig-Penney potential ($V_0=1$).
\label{fig:Bands}}
\end{center}
\end{figure*}

An example of the band structure for strong nonlinearity is shown
in Fig.~\ref{fig:Bands}(a). The interaction strength is chosen to
be a factor of ten larger than the strength of the potential,
causing the appearance of swallowtails. Only the lowest energy
band can be extracted with this class of solutions.  This shows
that these simple exact solutions are sufficient to describe
measurable properties of the condensate such as breakdown of
superfluidity for critical values of the
nonlinearity~\cite{Mueller2002,Fallani2004}.

In the work of Bronski et al.~\cite{Bronski2001,Bronski2001_2},
the linear stability properties of the solutions were proved
for the case of constant, or trivial phase. No such proof was
discovered for the case of non-trivial phase solutions. The latter
are, for example, the only stable ones in the exactly sinusoidal
case of $k=0$.  It was found numerically that non-trivial phase
solutions for repulsive condensates with large offsets $B$ were
stable, while for smaller $B$ they were unstable. No explanation
for this stability property was found. After we recast the
solutions in Bloch form and used them to determine the band
structure, the stability properties become immediately apparent.
Solutions on the upper edge of the swallowtail are known to be
unstable, as they represent an energy
maximum~\cite{Mueller2002,Machholm2003,Seaman2005_2}; these
correspond to small $B$. Solutions on the main part of the band
and the lower edge of the swallowtail represent an energy minimum
and are stable; these correspond to large $B$. For instance, in
Fig.~\ref{fig:Bands}(a), the main part of the band from $q=0$ to
$q=1$ corresponds to $B\in [0.559,0.509]$; the lower edge of the
swallowtail near the edge of the Brillouin zone corresponds to
$B\in[0.509,0.277]$; and the upper edge, which is unstable,
corresponds to $B\in[0.277,0]$.  We note that, in comparing with
Bronski {\it et al.}, they normalized the wavefunction to $gn$
rather than having $gn$ as the coefficient of the nonlinearity in
the NLS.  In their numerical studies of stability, they did not
hold their normalization (our nonlinear coefficient) fixed. In
Fig.~\ref{fig:Bands}(a), we fix the nonlinearity to be $gn=10
|V_0|$.

\subsection{Solution by Three Mode Approximation}
\label{subsec:Machholm}

In a subsequent work by Machholm, Pethick and
Smith~\cite{Machholm2003}, a more complete method to determine the
nonlinear band structure was introduced that used a sinusoidal
potential \be V(x)=2 V_0 \cos^2(\pi x)\, . \ee The wavefunction
was expanded in a Fourier series for a particular quasimomentum
and the energy was minimized. This method uses the exact physical
form of the typical experimental lattice potential that a BEC is
held in but requires the root finding of many free parameters.
Machholm {\it et al.} showed that one can obtain analytical
solutions by using a three-mode Fourier spectrum: \be
\Psi(x)=\sqrt{n} e^{i q x}(a_0+a_1e^{i2\pi x}+a_{-1}e^{-i2\pi
x})\, , \ee where $a_0$, $a_1$ and $a_{-1}$ are real coefficients.
Due to the normalization condition on the wavefunction, there are
two real free parameters of the solution.

Using the three-mode expansion of the wavefunction, it is possible
to extract the first two bands and the lower part of the third
band, as shown in Fig.~\ref{fig:Bands}(b).  Notice that, unlike
with the previous method, full information from the first two
bands can be extracted, instead of only the lowest band. However,
the three-mode approximation overestimates the width of the
swallowtails~\cite{Machholm2003}. If more Fourier components are
included, for a total between five and ten, the first three bands
can be described to within $1\%$ accuracy. For higher bands even
more Fourier components are needed. This method was later extended
to include period-doubled states~\cite{Machholm2004}.

In addition to determining the band structure, Machholm {\it et
al.} were able to determine several analytical expressions
associated with the stability properties of the condensate.  In
particular, the interaction strength at which swallowtails first
appear in the lowest band, and the width of the swallowtails for a
given interaction strength were determined analytically. Studies
of the stability of the condensate to small perturbations were
also performed.

\subsection{Solution via a Piece-wise Constant Lattice}
\label{subsec:Seaman}

In comparison, in our previous work~\cite{Seaman2005_2} a lattice
of delta functions, a Kronig-Penney potential, was used \be
V(x)=V_0 \sum_{j=-\infty}^{+\infty}\delta(x-j)\, . \ee  We
presented the complete set of Bloch waves solutions in analytical
form by solving the piece-wise-constant potential case and using
the appropriate boundary conditions to fix the parameters of the
solution.  The potential, however, is quite different from the
experimentally created sinusoidal lattice.  The optical lattice
potential is composed of a single Fourier component while the
Kronig-Penney potential is a comb of equally weighted Fourier
components.  We proved that the most general form of the density
over a finite interval of constant potential is given by \be
\rho(x)=B+\frac{k^2 b^2}{g}\sn^2(bx+x_0,k)\, , \ee where again
$\sn$ is one of the Jacobi elliptic functions.  It should be noted
that the square of any Jacobi elliptic function can be related
linearly with the square of every other Jacobi elliptic function.
Thus all twelve Jacobi elliptic functions are possible solutions.
The chemical potential $\mu$ and phase prefactor $\alpha$ (see
Eq.~(\ref{eqn:phase})) are given by \bea
\mu=\frac{1}{2}[b^2(1+k^2)+3Bg]\, ,
\\ \alpha^2=B(k^2b^2/g+B)(b^2+Bg)\, . \eea The energy bands can be
determined by varying one of the parameters in the wavefunction,
such as $B$ or $b$, and determining the quasimomentum and the
energy from Eqs.~(\ref{eqn:quasimomentum}) and~(\ref{eqn:energy}).
In particular, the parameter scaling $b$ was varied since it is
closely related to the quasimomentum.  The offset $B$ and elliptic
parameter $k$ are determined by number conservation and the
boundary conditions across the delta functions.

The first three bands for the Kronig-Penney lattice are presented
in Fig.~\ref{fig:Bands}(c).  Although only the first three bands
are presented, our method can be used to determine the higher
bands with no additional computational intensity, unlike with the
technique of Sec.~\ref{subsec:Machholm}.  For all energy bands,
all that is required computationally is the root finding of two
parameters, the offset $B$ and the elliptic parameter $k$.  The
stability properties of the condensate were then numerically
determined by dynamically evolving an initial state composed of
the stationary solution plus a small amount of white noise. It was
found that the stability of the energy bands depends on the
interaction strength. For weak nonlinearity, the bands become
unstable for quasimomentum greater than approximately $\pi/2$. For
strong nonlinearity, the tops of the swallowtails are unstable but
the remainder of the first band is stable.  The second and higher
bands are always unstable.

In spite of the differences in the potentials, all three methods
find a similar nonlinear band structure.  In addition, the
stability properties of the three models were also the same.
Therefore, to study the main physical aspects of this system, any
method can be used.  All three methods are useful in determining
the first band.  For the full band structure, only our method is
analytically and numerically tractable.

\section{The Two-color Lattice and Formation of Swallowtails}
\label{sec:TwoColorLattice}

\begin{figure} [tb]
\begin{center}
\epsfxsize=7.8cm \leavevmode \epsfbox{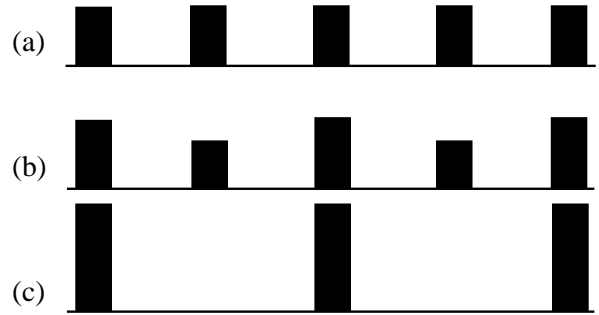} \caption{A
sketch of the potential is illustrated for the three cases of
Figs.~\ref{fig:TwoColor} and~\ref{fig:TwoColorStrong}, with
appropriately normalized boxes representing delta functions: (a) a
one-color lattice of period $d$ and $\Delta=0$; (b) a two-color
lattice with a small difference in the potential strengths, period
$2d$ and $\Delta=0.2$; (c) a one-color lattice with period $2d$
and $\Delta=1$. Adiabatically tuning the system from (a) to (c)
allows for a novel explanation of the appearance of swallowtails
in the band structure.\label{fig:deltas}}
\end{center}
\end{figure}

Current experiments with BEC's utilize only one-color lattices in
the weakly nonlinear regime.  Therefore there has been no
experimental observation of the swallowtail structure, which
requires strong nonlinearity. Two-color lattices, on the other
hand, allow one to produce swallowtails even for weak
nonlinearity, as we shall show in the following. Moreover,
adiabatically tuning the two-color lattice allows for an
explanation of swallowtails for both repulsive and attractive
BEC's.

A two-color lattice is produced by adding a second frequency
component of twice the fundamental frequency, or half the period.
Such a straightforward modification of existing experiments can be
made via second harmonic generation with nonlinear crystals.  In
Sec.~\ref{sec:Review} we showed that the exact form of the
potential is unimportant.  Therefore, to study the two-color
lattice, we use a Kronig-Penney-like potential with delta
functions of two different strengths: \be
V(x)=V_0\sum_{j=-\infty}^{+\infty}[(1-\Delta)\delta(x-2
j)+(1+\Delta)\delta(x-2 j+1)]\, , \label{eqn:twocolorpot}\ee where
$\Delta\in[0,1]$.  The strengths of the two sublattices are
$(1-\Delta)V_0$ and $(1+\Delta)V_0$, respectively.  The form of
Eq.~(\ref{eqn:twocolorpot}) allows the potential to be tuned
continuously from the one-color lattice of strength $V_0$ and
period $d$ ($\Delta=0$) to the one-color lattice of strength
$2V_0$ and period $2d$ ($\Delta=1$). This holds the average
potential strength constant. For $0<\Delta<1$ the potential is
two-color.  Figure~\ref{fig:deltas} sketches the progression
between the two lattices, where the delta functions are
represented by square functions of fixed width and varying
heights.  We will show how period-doubled solutions of the
$\Delta=0$ lattice map onto Bloch-wave solutions of the $\Delta=1$
lattice, including the swallowtails.

\begin{figure}
\begin{center}
\epsfxsize=7.8cm \epsfysize=17cm \leavevmode
\epsfbox{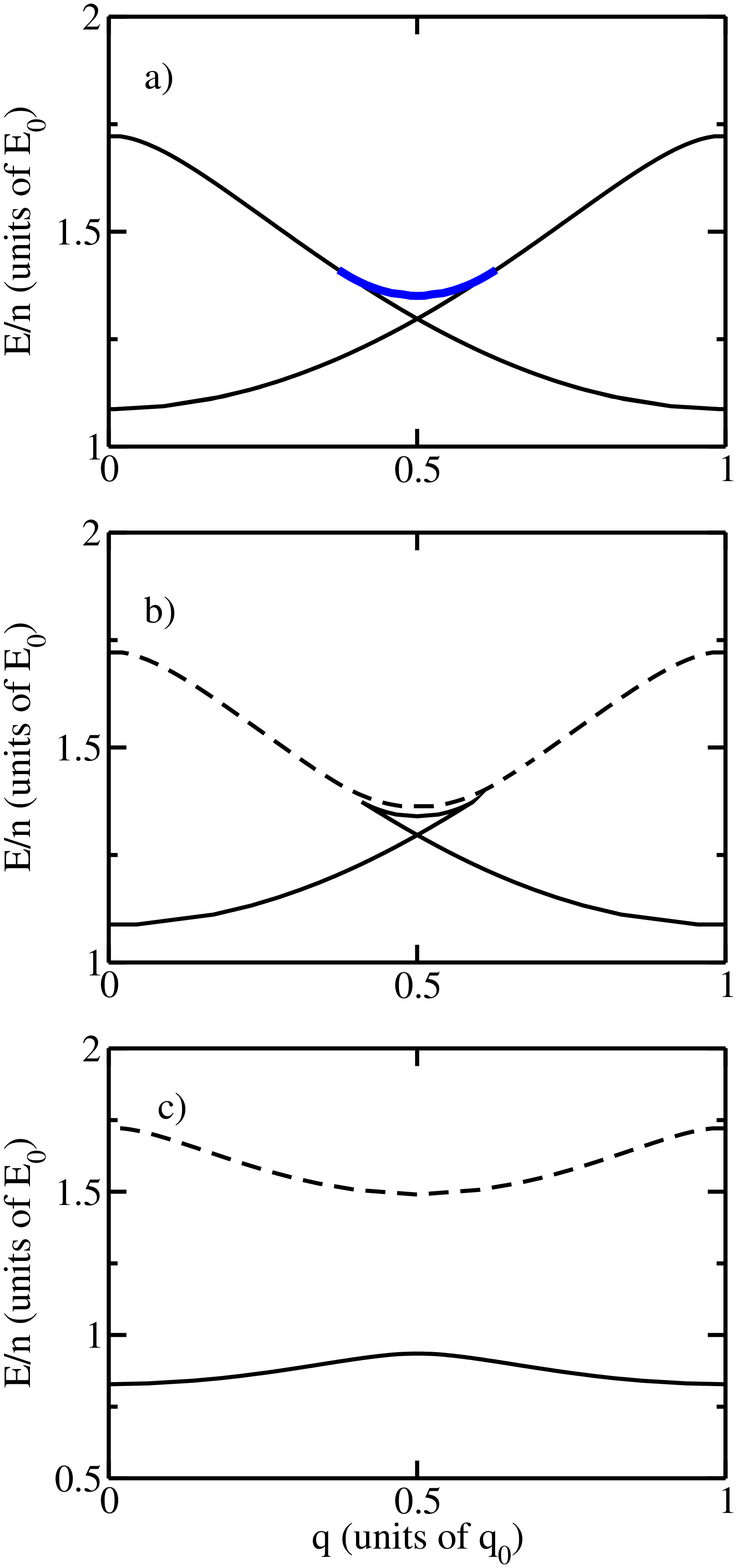} \caption{(color online) Band
structure for a two-color lattice with weak nonlinearity,
$gn=V_0$. (a) Shown are period-doubled solutions for a one-color
lattice of period $d$ and $\Delta=0$ (Thin black curve: trivial
solutions; thick blue curve: non-trivial); (b),(c) Bloch-wave
solutions with period $2d$ (Solid curve: lowest band; dashed
curve: second band). (b) A two-color lattice with $\Delta\ll 1$.
Note that the central swallowtail is derived from the non-trivial
period-doubled solution of panel (a). (c) A one-color lattice with
$\Delta=1$. The swallowtail has disappeared due to the weak
nonlinearity. The lattices associated with (a)-(c) are sketched in
Fig.~\ref{fig:deltas}. \label{fig:TwoColor}}
\end{center}
\end{figure}

The stationary states of the NLS with the potential of
Eq.~(\ref{eqn:twocolorpot}) are determined in the same manner as
was described in Sec.~\ref{subsec:Seaman}, with the addition of a
second boundary condition.  Period-doubled solutions can be
obtained via the transformation $\Psi(x)=f_q(x)\exp(i 2 q x)$,
while Bloch solutions were found via $\Psi(x)=f_q(x)\exp(i q x)$.
In general, one can obtain solutions of the form
$\Psi(x)=f_q(x)\exp(i m q x)$, with m an integer; we focus on the
period-doubled case, $m=2$, and the Bloch wave case, $m=1$.

\begin{figure}[tb]
\begin{center}
\epsfxsize=7.8cm \leavevmode \epsfbox{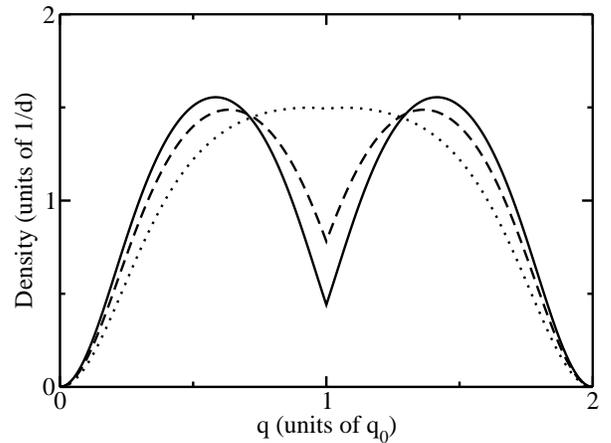}
\caption{The condensate density profiles are illustrated for three
cases: $\Delta=0$, non-trivial period-doubled solutions (solid
curve); $\Delta = 1/2$, upper edge of the swallowtail (dashed
curve); and $\Delta = 1$, the lowest band (dotted curve).  The
parameters are $q=0.5$ and $gn=V_0$, which is the regime of weak
nonlinearity, as in Fig.~\ref{fig:TwoColor}. \label{fig:density}}
\end{center}
\end{figure}

We first illustrate period-doubled solutions for the lattice of
Fig.~\ref{fig:density}(a), $\Delta=0$.  There are two types of
period-doubled solutions, \emph{trivial} and \emph{non-trivial}.
The trivial period-doubled solutions simply reproduce the Bloch
waves. That is to say, solutions of period $d$ are also trivially
solutions of period $2d$.  These are illustrated in
Fig.~\ref{fig:TwoColor}(a) as thin curves for the case of weak
nonlinearity. Period-doubled solutions extend across a Brillouin
zone of half the quasimomentum-domain as that of Bloch waves. Thus
the domain of Fig.~\ref{fig:TwoColor} ($q\in[0,1]$) is half that
of Fig.~\ref{fig:Bands} ($q\in[-1,1]$). The bands are required to
be symmetric around the center of the Brillouin zone. One reflects
the band in the right half of the Bloch Brillouin zone around its
center at $q=0.5$ to obtain the trivial period-doubled solutions.
This leads to the two branches which make the form of an `x' in
Fig.~\ref{fig:TwoColor}(a) (thin black curves). The non-trivial
period-doubled solutions are shown as a thick blue curve in
Fig.~\ref{fig:TwoColor}(a).  These form a saddle between the
trivial solution branches.  The form of the density for the
non-trivial period-doubled solutions is shown as the solid curve
in Fig.~\ref{fig:density}.  The non-trivial solutions are caused
by the nonlinearity in Eq.~(\ref{eqn:NLS}); they have no analog in
the linear Schr\"odinger equation.  Comparing
Fig.~\ref{fig:TwoColor} to the swallowtails illustrated in
Fig.~\ref{fig:Bands}, we make the following conjecture:
non-trivial period-doubled solutions for a lattice of period $d$
appear as the saddle of the swallowtail for Bloch waves for a
lattice of period $2d$.

This conjecture is supported by tuning $\Delta$ from zero to
unity.  We illustrate this tuning in the subsequent panels of
Fig.~\ref{fig:TwoColor}. Figure~\ref{fig:TwoColor}(b) shows the
case $\Delta\ll 1$. Figure~\ref{fig:TwoColor}(c) shows the
endpoint, $\Delta=1$.  The shape of the lattice in the panels
(a)-(c) is sketched in Fig.~\ref{fig:deltas}. In
Fig.~\ref{fig:TwoColor}(b), the trivial and non-trivial
period-doubled solutions of Fig.~\ref{fig:TwoColor}(a) separate
into two bands. The lower band (solid curve) consists of the lower
part of the trivial period-doubled solutions, with the non-trivial
period-doubled solutions forming the upper edge of the
swallowtail. The upper band (dashed curve) consists of the upper
part of the trivial period-doubled solutions.  One observes that
there is a small gap between the two bands.  Moreover, unlike in
the case of the one-color lattice, one obtains a swallowtail even
for weak nonlinearity.  As $\Delta$ is increased, this gap
increases. Figure~\ref{fig:TwoColor}(c) shows the case $\Delta=1$.
The system is now again in a weakly nonlinear regime with a
one-color lattice.  Thus the swallowtail disappears.

\begin{figure} [tb]
\begin{center}
\epsfxsize=7.8cm \epsfysize=17.5cm \leavevmode
\epsfbox{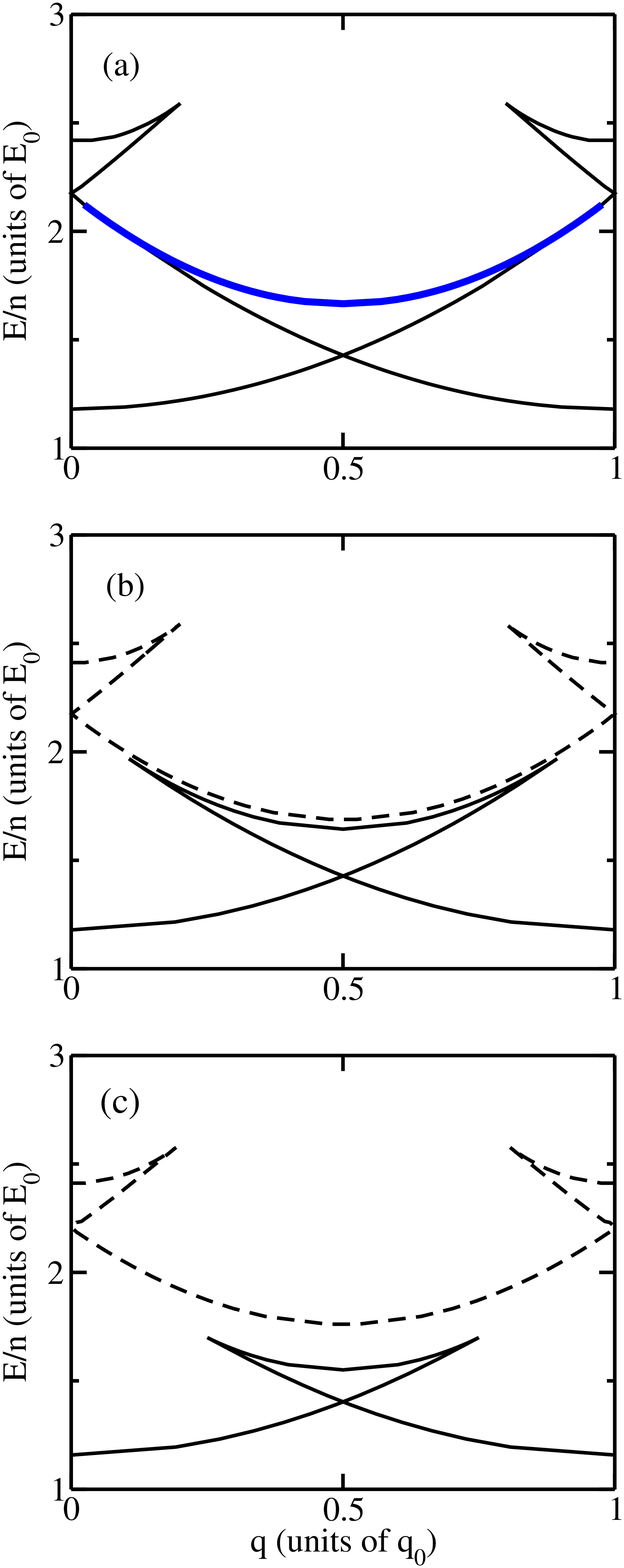} \caption{(color online) Same as
Fig.~\ref{fig:TwoColor} but for strong nonlinearity, $gn=10V_0$.
In (b), (c) note that the upper edge of the central swallowtail in
the lower band is derived from the non-trivial period-doubled
solution of panel (a) (blue thick curve). The lattices associated
with (a)-(c) are sketched in
Fig.~\ref{fig:deltas}.\label{fig:TwoColorStrong}}
\end{center}
\end{figure}

One observes that, in general, the first two Bloch-wave bands of a
one-color lattice of period $2d$ (panel (c)) can be derived from
the trivial period-doubled solutions of a lattice of period $d$
(panel (a)) via a two-color intermediate lattice (panel (b)). This
follows from the fact that the second Bloch-wave band has two
density peaks per site.

In the strongly nonlinear regime, the swallowtail forms in a
similar manner but does not disappear for $\Delta=1$.  We
illustrate this sequence in Fig.~\ref{fig:TwoColorStrong}, again
for $\Delta=0$, $\Delta\ll 1$, and $\Delta=1$.  One clearly sees
that the upper edge of the swallowtail in the one-color lattice of
period $2d$ is derived from the non-trivial period-doubled
solutions of the one-color lattice of period $d$.  The lower edge
of the swallowtail is derived from the trivial period-doubled
solutions.

This introduces a novel perspective on the origin of the
swallowtails in the nonlinear band structure.  In the figures we
illustrated the case of a repulsive condensate, as that is the
most common experimentally. The same argument holds for attractive
condensates, where the swallowtails form below, rather than above
the bands.  In particular, in our study of the one-color
Kronig-Penney potential~\cite{Seaman2005_2}, we found it
intriguing that no swallowtail formed on the lowest band for the
attractive case.

Previous interpretations of swallowtails were based on the
superfluid screening properties of the
condensate~\cite{Mueller2002}.  In this argument, the condensate
sees the quadratic free particle dispersion up to the sound speed.
When written in the form of the Bloch ansatz, these quadratic
curves repeat in each Brillouin zone.  When the sound speed is
such that the maximum quasimomentum of each curve overlaps with
the curve from the adjoining Brillouin zone, one obtains
swallowtails. This does not apply to attractive condensates, since
the swallowtails form on the lower edge of the bands and not at
all on the lowest band.

In contrast, our argument based on period-doubling applies to both
repulsive and attractive condensates.  For attractive
nonlinearity, the non-trivial period-doubled states form a convex
saddle on the \emph{lower} edge of the `x' of
Fig.~\ref{fig:TwoColor}(a), rather than the upper edge.  As
$\Delta$ is tuned from zero to unity, the upper part of the `x'
separates from the lower part, carrying the swallowtail with it.
Thus no swallowtail can form on the lowest band.

Thus, we understand swallowtails to be adiabatically connected to
period-doubled solutions of a lattice of half the period.  They
originate in non-trivial period-doubling brought about by the
nonlinear term in the NLS.

Finally, since the density is a key observable in experiments on
BEC's, in Fig.~\ref{fig:density} density profiles are illustrated
for weak nonlinearity and $\Delta=0,\,1/2,\,1$.  The plot is made
for quasimomentum $q=0.5$ and the energy per particle such that
the solution lies at the center of the non-trivial period-doubled
solution (thick blue curve) in Fig.~\ref{fig:TwoColor}(a), at the
top of the central swallowtails on the lowest band in
Fig.~\ref{fig:TwoColor}(b), and on the lower band in
Fig.~\ref{fig:TwoColor}(c).

\section{Stability Properties and Comparison with Experiments}
\label{sec:Stability}

Several important experiments have been performed with superfluid
BEC's in optical lattices.  In particular, an experiment by
Fallani {\it et al.}~\cite{Fallani2003,deSarlo2004} studied the
loss rate of the BEC as the quasimomentum was varied. This loss
rate is expected to be monotonically related to the instability
time of the condensate.  This observable can be modelled
theoretically by adding white noise to an initial stationary state
on the lattice in simulations.  In this way, we investigate the
nonlinear stability properties of Bloch-wave solutions to the
two-color lattice.  Note that unstable solutions which have
lifetimes longer than experimental timescales will appear
experimentally stable.

In our simulations, Eq.~(\ref{eqn:NLS}) was dynamically evolved
using a variable step fourth-order Runga-Kutta algorithm in time
and a pseudospectral method in space.  We chose periodic boundary
conditions in one dimension, with a sufficient number of sites so
that the results were independent of the ring circumference.  Two
schemes for delta functions on a grid were considered: point
defects, and narrow square potentials covering several grid
points. For square potentials less than one tenth the lattice
period $d$, differences between the two schemes were negligible.
Simulations were performed over time scales on the order of those
relevant to the experiments, namely, hundreds of milliseconds. The
lattice spacing $d$ is taken to be $1$~$\mu$m and the atomic mass
to be that of $^{87}Rb$.

\begin{figure}[tb]
\begin{center}
\epsfxsize=7.8cm \leavevmode \epsfbox{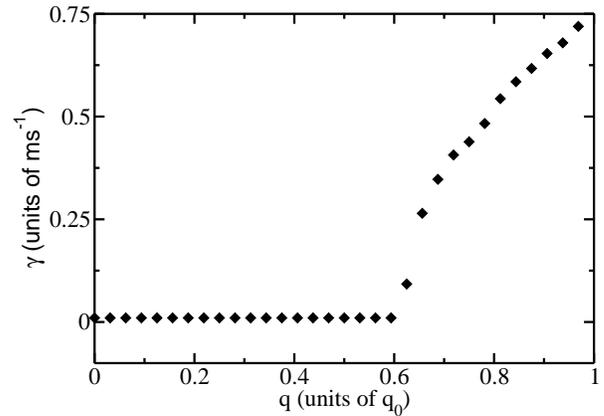}
\caption{Shown is the inverse of the instability time
$\gamma=T_i^{-1}$ of stationary states in the lowest band, as a
function of the quasimomentum.  White noise was added to the
initial state and the wavefunction was evolved numerically for the
NLS with a one-color lattice and weak nonlinearity, $gn=V_0$.
\label{fig:stabTimes}}
\end{center}
\end{figure}

The stability of the system is quantified by the variance in the
Fourier spectrum, \be \sigma(t) \equiv
\frac{\sum(f(p,t)-f(p,0))^2}{2\sum(f(p,0))^2}\, , \ee where
$f(p,t)$ is the Fourier component of the wavefunction at momentum
$p$ and time $t$.  The variance approaches unity when the
wavefunction has no component in common with the initial state and
zero when then the wavefunction it is identical to the initial
state. We define the instability time $T_i$ to occur
when the variance has increased to $\sigma=0.5$.

A clear experimental observable is the rate of loss of atoms from
the system, since the density of the sample can be imaged at
successive times. This loss rate is expected to be monotonically
related to the inverse time in which instabilities arise.
Therefore, we present two relevant studies of the instability
times of the condensate. First, the growth rate of the
instability, \be \gamma\equiv 1/T_i\,,\ee is determined for a
one-color lattice and weak nonlinearity.  This relates directly to
the experiment of Fallani {\it et al.}.  Second, the lowest
positive quasimomentum for which the system becomes unstable is
evaluated as a function of $\Delta$ for the two-color lattice.

Figure~\ref{fig:stabTimes} presents the results of the first
study.  The solutions are stable until the quasimomentum where the
nontrivial and trivial period-doubled solutions connect in the
band structure (see Fig.~\ref{fig:TwoColor}(a), as well as
Ref.~\cite{Machholm2004}). Stability can also be determined from
the effective mass of the system, \be
m^*\equiv\frac{1}{\partial^2E/\partial^2q}\, . \ee If the
effective mass becomes negative, the NLS becomes effectively
attractive.  Since the ground state of an attractive condensate is
localized, solutions with negative effective mass $m^*<0$ are
unstable.  The point where the nontrivial and trivial
period-doubled solutions connect corresponds to a change in the
sign of the effective mass from positive to negative. Our figure
qualitatively replicates the results found by Fallani {\it et
al.}~\cite{Fallani2004} (see figures therein~\cite{Fallani2004}).

\begin{figure}[tb]
\begin{center}
\epsfxsize=7.8cm \leavevmode \epsfbox{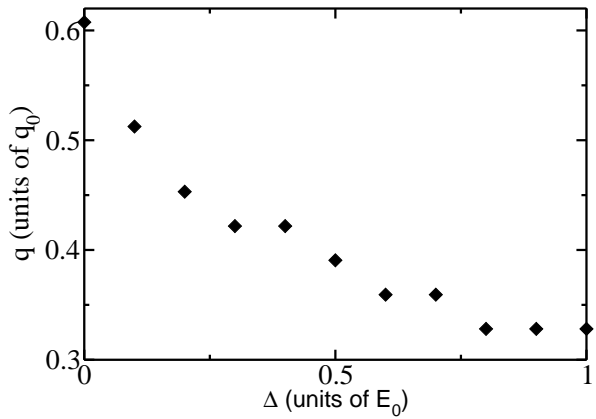} \caption{The
lowest positive quasimomentum for which the system becomes
unstable is determined as a function of $\Delta$ as the one-color
lattice of period $d$ is tuned via a two-color lattice to the
one-color lattice of period $2d$. The result of simulations of the
NLS with weak nonlinearity, $gn=V_0$, are shown.
\label{fig:stabQuasi}}
\end{center}
\end{figure}

Figure~\ref{fig:stabQuasi} presents the second study.  In
particular, the smallest quasimomentum for which the periodic
system first becomes unstable in a time less than $100$ ms is
plotted.  The two end points at $\Delta=0$ and $\Delta=1$ are
easily determined, since those systems are one-color lattices.  As
per our discussion in the previous paragraph, the Bloch wave
solutions become unstable at the point where the non-trivial
period-doubled solutions connect to the Bloch-wave band.  The two
one-color lattice systems have the same ratio of nonlinearity to
potential strength, since for $\Delta=0$ the potential strength is
$V_0$ and the number of atoms is $n$, while for $\Delta=1$ they
are $2V_0$ and $2n$.  Thus the existence of the non-trivial
period-doubled states scales with the length of the Brillioun
zone. The absolute quasimomentum where instability occurs then
differs by a factor of two, since for a lattice of period $2d$ the
Brillouin zone is half the length as for $d$.  In
Fig.~\ref{fig:stabQuasi}, this is $q\sim 0.6$ for $\Delta=0$ and
$q\sim 0.3$ for $\Delta=1$.  In the case of the linear
Schr\"odinger equation, the endpoints would be $0.5$ and $0.25$,
respectively.  Note that our simulations are accurate to within a
few percent.

The intermediate points in Fig.~\ref{fig:stabQuasi}, where
$0<\Delta<1$ and one obtains a two-color lattice, can be
understood by again considering where the effective mass becomes
negative.  As the energy band from the $\Delta=0$ lattice
separates into the first two bands of the $\Delta=1$ lattice, the
upper edge of the swallowtail shrinks.  Recall that states on the
lower edge of swallowtails are stable~\cite{Seaman2005_2}. As
$\Delta$ is increased, the swallowtail disappears, since the
nonlinearity is weak. The quasimomentum at which the band changes
from concave up to concave down, i.e. where the effective mass
changes sign, moves from right to left, as can be seen in the
descending data points of Fig.~\ref{fig:stabQuasi}.  A similar
argument holds for the case of strong nonlinearity.

\section{Conclusion} \label{sec:Conclusion}

We have presented a novel way of understanding the appearance of
swallowtails in nonlinear band structure.  Swallowtails in a
lattice of period $2d$ are adiabatically connected to the
non-trivial period-doubled solutions of a lattice of period $d$.
By non-trivial we mean solutions not immediately obvious from the
fact that a solution with period $d$ is also a solution of period
$2d$.  Non-trivial period-doubled solutions are caused by the
nonlinearity in the mean field of the
condensate~\cite{Machholm2004}.  They do not appear in the linear
Schr\"odinger equation and therefore have no analog in linear band
theory. Our way of understanding swallowtails is valid for both
repulsive and attractive condensates, unlike previous
explanations~\cite{Mueller2002}.

We used the two-color lattice to adiabatically connect the
one-color lattice of period $d$ and $2d$.  We showed that
swallowtails appear in the band structure of the two-color lattice
even for weak nonlinearity, unlike in the one-color case.  We then
made explicit predictions for the onset of instability in the band
structure for the one- and two-color lattices based on numerical
simulations. Instability is experimentally observable as an
increased loss rate due to heating of the
condensate~\cite{Fallani2004}, and weak nonlinearity is the
present regime of experimental investigation.

Our studies utilized a two-color Kronig-Penney potential.  In
order to justify the use of this model we compared analytic
solutions to the one-color Kronig-Penney~\cite{Seaman2005_2},
Jacobi elliptic~\cite{Bronski2001}, and sinusoidal
potentials~\cite{Machholm2003}.  We showed that each of these
models resulted in a similar band structure. Moreover, by putting
the exact solutions to the Jacobi elliptic potential in Bloch wave
form, we were able to resolve a longstanding question concerning
density offsets and stability in the work of Bronski {\it et
al}~\cite{Bronski2001,Bronski2001_2}.

We note that there are other solution types than the ones we have
considered. For instance, there are gap
solitons~\cite{Louis2003,Louis2005,Merhasin2005} and time-periodic
solutions~\cite{Porter2004,Porter2004_2}.  These can have novel
features in the two-color lattice~\cite{Louis2005}. We emphasize
that we have treated only the superfluid phase of cold bosons on a
lattice~\cite{Fisher1989,Greiner2002}.

\begin{center}
{\bf Acknowledgments}
\end{center}

We thank Eugene Zaremba and Christopher Pethick for useful
discussions. Support is acknowledged for B.T.S. and M.J.H. from
the National Science Foundation and for L.D.C. from the U.S.
Department of Energy, Office of Basic Energy Sciences via the
Chemical Sciences, Geosciences and Biosciences Division.

\bibliographystyle{prsty}

\end{document}